\documentclass[pra,showpacs,superscriptaddress,reprint,amsmath,amssymb,aps,nofootinbib]{revtex4-1}
\usepackage{graphicx}
\usepackage{bm}
\usepackage{epstopdf}
\usepackage{siunitx}
\usepackage{color}

\DeclareSIUnit\gauss{G}

\usepackage[
pdfstartview=FitH,
bookmarks=true,
bookmarksopen=true,
breaklinks=true,
colorlinks=true,
linkcolor=blue,anchorcolor=blue,
citecolor=blue,filecolor=blue,
menucolor=blue,pagecolor=blue,
urlcolor=blue]{hyperref}
\usepackage{float}
\usepackage[toc,page,title,titletoc,header]{appendix}

\begin{document}

\title{Implementation of a double-path multimode interferometer using a spinor Bose-Einstein condensate}

\author{Pengju Tang}
\affiliation{State Key Laboratory of Advanced Optical Communication System and Network, Department of Electronics, Peking University, Beijing 100871, China}
\author{Xiangyu Dong}
\affiliation{State Key Laboratory of Advanced Optical Communication System and Network, Department of Electronics, Peking University, Beijing 100871, China}
\author{Wenjun Zhang}
\affiliation{State Key Laboratory of Advanced Optical Communication System and Network, Department of Electronics, Peking University, Beijing 100871, China}
\author{Yunhong Li}
\affiliation{School of Electronics and information, Xi'an Polytechnic University, Xi'an 710048, China}
\author{Xuzong Chen}
\affiliation{State Key Laboratory of Advanced Optical Communication System and Network, Department of Electronics, Peking University, Beijing 100871, China}
\author{Xiaoji Zhou}\email{xjzhou@pku.edu.cn}
\affiliation{State Key Laboratory of Advanced Optical Communication System and Network, Department of Electronics, Peking University, Beijing 100871, China}
\affiliation{Collaborative Innovation Center of Extreme Optics, Shanxi University, Taiyuan, Shanxi 030006, China}

\date{\today}

\begin{abstract}

We realize a double-path multimode matter wave interferometer with spinor Bose-Einstein condensate and observe clear spatial interference fringes as well as a periodic change of the visibility in the time domain, which we refer to as the time domain interference and which is different from the traditional double-path interferometer. By changing the relative phase of the two paths, we find that the spatial fringes first lose coherence and recover. As the number of modes increases, the time domain interference signal with the narrower peaks is observed, which is beneficial to the improvement of the resolution of the phase measurement. We also investigated the influence of initial phase configuration and phase evolution rate between different modes in the two paths. With enhanced resolution, the sensitivity of interferometric measurements of physical observables can also be improved by properly assigning measurable quantities to the relative phase between two paths.

\end{abstract}


\maketitle

\section{Introduction}

Matter wave interferometers have brought us exciting opportunities of implementing high-precision measurements in both science and technology, including quantum information, quantum simulation, and gravity measurements\cite{3cronin2009optics,2Kurn2013spinor,3.1HMuntinga2013Microgravity, 3.0lombardi2014reading}, as well as insight into fundamental issues of quantum physics such as internal properties of atoms, many-body phenomena, and the interplay between general relativity and quantum mechanics\cite{1dalfovo1999theory,6Simsarian2000phase,3.5wheeler2004spontaneous,3.2Dimopoulos2007Relativity,3.6grond2010atom, 3.4Pikovski2017clock}. Much research focused on achieving a high performance of interferometers has been carried out\cite{4Schumm2005phase,4.1PhysRevAYANG,5Ramsey}. In particular, sensitivity, one of the important indicators of the ability of an interferometer, received much attention. Most of the previous research implemented a double-path configuration as the model of studying the sensitivity of measurements and gave the limits of the sensitivity, shot-noise limit, and sub-shot-noise limit\cite{5.1PhysRevLett.102.100401,5.2Lucke773}. However, in experiment it requires a lot of effort to reduce the noise to reach the noise limit.

Therefore, to suppress the noises from experimental systems and improve the resolution (and thus the sensitivity) of the signal becomes a major concern. On the one hand, a noise-resilient parallel multicomponent interferometer\cite{7.1PhysRevATPJ} was implemented and effectively enhanced the robustness of measurements of phases. On the other hand, some proposed to increase the number of paths $M$ so as to sharpen the peaks in the interference fringes\cite{7.3three-path,7.4weitz1996multiple,7.2multiphoton,7.5PhysRevA.87.033607,8paul2017measuring}. It has been demonstrated by experiments of internal-state multipath interferometers using spinor Bose-Einstein condensates (BECs) that, as long as the scaling of slope with $M$ exceeds $\sqrt{M}$ scaling of the shot noise, the sensitivity improves with the number of paths\cite{7.4weitz1996multiple,10petrovic2013multi}.

Further, some people sought to implement double-path multimode interferometers\cite{10.3PhysRevLett.88.100401,12.1Zych2011clock,10.1PhysRevLett.108.130402,10.2PhysRevLett.113.023003} to improve the resolution, which means there are several components (e.g., different spin orientations of a BEC) in each path instead of a single component in traditional interferometers. Its advantage is that additional modes would not boost the shot noise, as long as the modes do not interact with each other\cite{14PhysRevA.67.053803,11PhysRevA2017multimode}. The modes of a multimode interferometer could be chosen as either spin states or noninteracting external states. Especially, double-path two-mode interferometers\cite{12margalit2015self} (using spin states as the modes)  have shown that the modes in each path exhibit good controllability of phase under magnetic fields. 
These experiments, together with one of our previous works\cite{13tang2019implementation} which implemented a full-spin Stern-Gerlach interferometer, demonstrated the feasibility of using internal states to implement multimode interferometers.

In this paper, we propose and realize a method to implement a double-path multimode interferometer with spinor BECs as internal modes to investigate the resolution by using the techniques of Majorana transition and radio-frequency (rf) coupling. In a gradient magnetic field, the phase evolution rate of the two paths has a tunable fixed difference, which allows us to control the relative phase between modes by tuning the phase evolution time. Therefore, unlike traditional double-path interferometers, our interferometer not only shows spatial interference fringes, but also exhibits a periodic dependence of visibility on the phase evolution time, which we refer to as the time domain interference. This so-called time domain fringe could be used to measure the phase shift accumulated in the evolution of each mode, which we refer to as the multimode stage below, and the corresponding observables. In general, it can be replaced by any physical quantity that can induce a difference between the phase evolution rate of different modes. The sharpness of the time domain fringes improves as the number of modes increases, which allows us to enhance the resolution of the interferometer. Our experiments demonstrate the existence of the time domain fringes and its dependence on number of modes and other experimental parameters. The results show that the resolution of phase measurements is increased to nearly twice compared with traditional double-path two-mode interferometers.

The remainder of this paper is organized as follows. In Sec.~\ref{sec:theory}, we introduce general procedures of a double-path multimode interferometer and give a brief theoretical analysis of the spatial interference. In Sec.~\ref{sec:experiment}, we describe our experiment setup and demonstrate the spatial interference pattern experimentally. The dependence of the time domain fringes on number of modes and other experimental parameters are demonstrated both theoretically and experimentally in Sec.~\ref{sec:time domain fringe}. Then we discuss the robustness of the measurements in Sec.~\ref{sec:discussion} and give a conclusion in Sec.\ref{sec:conclusion}.

\section{Principles of the double-path multimode interferometer}\label{sec:theory}

\subsection{Procedures of the double-path multimode interferometer}

The essential components and procedures of a double-path multimode interferometer are shown in Fig.~\ref{fig:procedure}. The whole procedure can be divided into three steps. 

\subsubsection{\textit{\textbf{Step 1: Coherent Stern-Gerlach momentum splitting}}}
A BEC on single magnetic sublevel $\left\vert F,m_{F}\right\rangle$ is prepared in a harmonic trap as the initial state [Fig.~\ref{fig:procedure}(a)], where $F$ and $m_F$ are the hyperfine structure quantum number and magnetic quantum number, respectively. The harmonic trap remains existing until the beginning of a time of flight (TOF). Through a Majorana transition, the internal state of the atoms is transferred into a superposition state of two magnetic sublevels $\left\vert F,m_{F}^{(\textit{I})}\right\rangle$ and $\left\vert F,m_{F}^{(\textit{II})}\right\rangle$ with equal coefficients (we denote these sublevels as $\left\vert \textit{I}\right\rangle$ and $\left\vert \textit{II}\right\rangle$ and leave out the hyperfine quantum number $F$ for convenience in the following discussions). The double-path configuration is achieved by coupling external states (i.e., positions and momenta) and internal states together [Fig.~\ref{fig:procedure}(b)] through a gradient magnetic field which lasts for $T_d$ ($\textit{I}$ and $\textit{II}$ also denote the two paths in the following). 

\subsubsection{\textit{\textbf{Step 2: Splitting the wave packet in each path into N modes and the evolution in the path}}}
A rf pulse is implemented to further separate each state $\left\vert \textit{I}\right\rangle$ ($\left\vert \textit{II}\right\rangle$) into $N=2F+1$ sublevels, followed by an evolution of time $T_N$ in a gradient magnetic field, constructing the multimode structure [Fig.~\ref{fig:procedure}(c)]. 

\subsubsection{\textit{\textbf{Step 3: Interference and observation}}}
\begin{figure}[htp]
	\includegraphics[width=\linewidth]{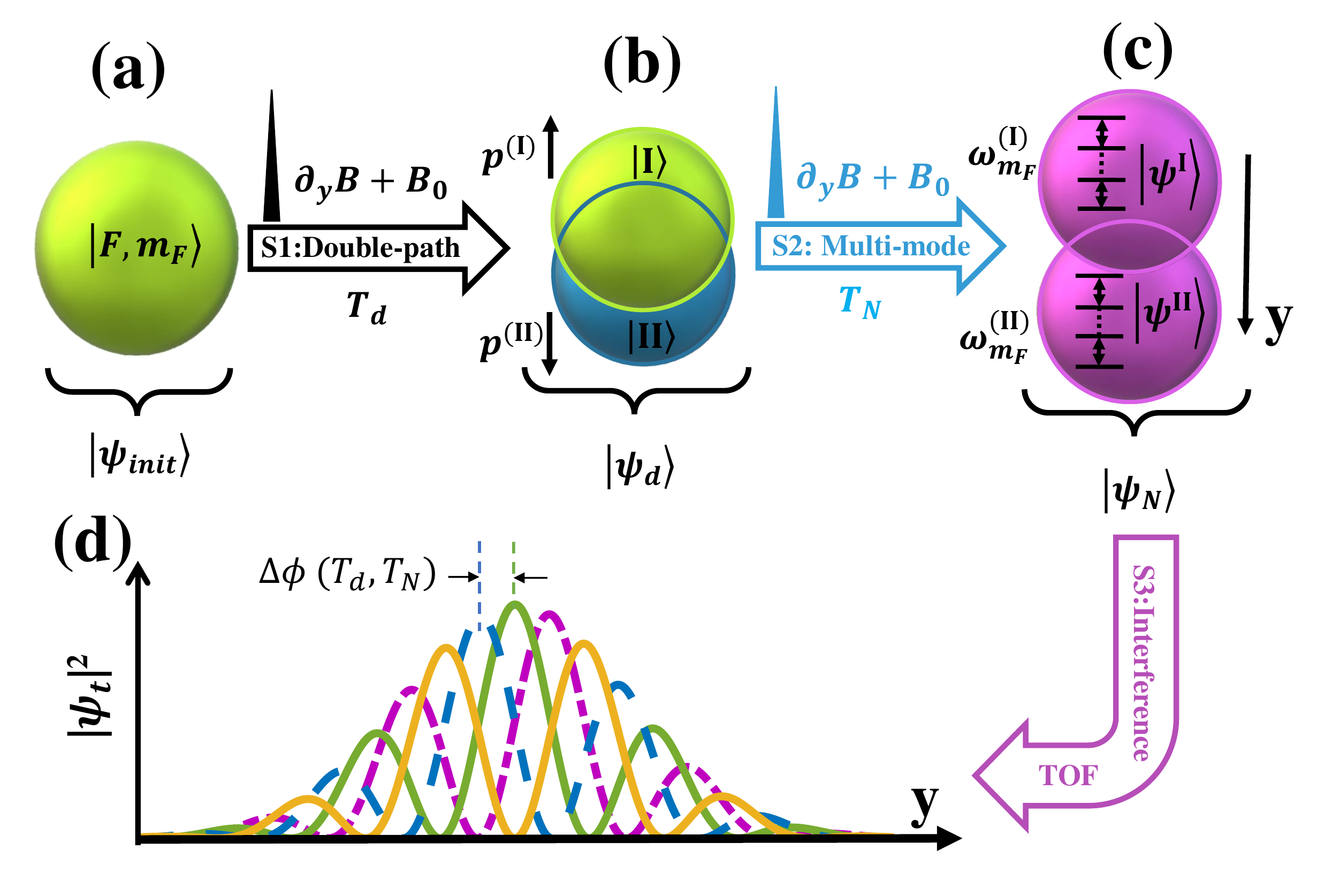}
	\caption{(Color online) The schematic of a double-path multimode  interferometer. An initial BEC in panel (a) is converted to two coherent wave packets serving as paths $|\textit{I}\rangle$ and $|\textit{II}\rangle$ in panel (b) (the momenta of which are $p^{(\textit{I})}$ and $p^{(\textit{II})}$, respectively) through a Majorana transition and an evolution of time $T_d$ in the gradient magnetic field, which we refer to as the step 1 (S1), or the double-path operation. In our Step 2(S2), each wave packet is converted to a superposition of several Zeeman sublevels by a rf pulse. After the evolution in the gradient magnetic field with different phase evolution rate $\omega_{m_F}^{(\textit{I}, \textit{II})}$ during time $T_N$, we got the superposition states in panel (c). Finally, after the time of flight (TOF) we observe the interference (d) of two superposition states by absorption imaging, where $\Delta\phi(T_d,T_N)$ is the relative phase between adjacent mode fringes. The fringe in each color represents the interference between the two wave packets of a single mode.}
	\label{fig:procedure}
\end{figure}

Finally, the harmonic trap is suddenly switched off. During the TOF, the atomic clouds expand and interfere with each other [Fig.~\ref{fig:procedure}(d)], before an absorption imaging is taken. 

\subsection{Analysis of the spatial interference}\label{subsec:InterferenceAnalysis}

In this subsection, we analyze the spatial interference of the double-path multimode interferometer and give the wave function for each stage shown in Fig.~\ref{fig:procedure}. Each state can be expressed as a direct-product of internal state, which takes into account the multimode effect and external states.

\subsubsection{\textit{\textbf{Initial state $\left|\psi_{\textit{init}}\right\rangle$ [Fig.~\ref{fig:procedure}(a)]}}}
The initial BEC can be denoted as $\left|\psi_{\textit{init}}\right\rangle= \left\vert F, m_F\right\rangle \left\vert {p_0, y_0}\right\rangle$ (we confine our description of external states to the $y$ direction, which is the direction of gravity, for simplicity). Here in the Dirac notation for external states, $p_0$ and $y_0$ are the momentum and the position of the center of mass of the wave packets, the same as below.

\subsubsection{\textit{\textbf{Double-path state $\left|\psi_{d}\right\rangle$ [Fig.~\ref{fig:procedure}(b)]}}}
After the double-path operation $T_d$, the state becomes
\begin{align}\label{eq:eq1}
	\left|\psi_{d}\right\rangle= \frac{1}{\sqrt{2}} \sum_{j=\textit{I},\textit{II}} e^{-i\xi^{(j)}}\left\vert j\right\rangle \left\vert {p^{(j)}, y^{(j)}}\right\rangle 
\end{align}
Here, the phases $\xi^{(\textit{I}, \textit{II})}$ are due to the state-dependent (and thus path-dependent) evolution. In the magnetic field, the phase evolution rate of each state is determined by its local energy, which is a sum of the kinetic energy and the Zeeman-split energy\cite{13machluf2013coherent,12margalit2015self}. Therefore, the momentum, the magnetic sublevel, and the local amplitude of the magnetic field together determine the phase evolution rate. 

\subsubsection{\textit{\textbf{Double-path multimode state $\left|\psi_{N}\right\rangle$ [Fig.~\ref{fig:procedure}(c)]}}}
In the next step to implement the multimode, both $\left\vert \textit{I}\right\rangle$ and $\left\vert \textit{II}\right\rangle$ are further separated into $N$ sublevels by the rf pulse, respectively. After the second phase evolution $T_N$ in two paths, the state becomes
\begin{align}
	\left|\psi_{N}\right\rangle &= \frac{1}{\sqrt{2}} \sum_{j=\textit{I},\textit{II}} \sum_{m_F} b_{m_F}^{(j)} \left|m_F\right\rangle e^{-i[\theta_{m_F}^{(j)}+\varphi_{m_F}^{(j)}]} \left\vert {p^{(j)},y^{(j)}}\right\rangle 
\end{align}
where $\theta_{m_F}^{(\textit{I},\textit{II})}$ is the initial phase from the rf pulse, $\varphi_{m_F}^{(\textit{I},\textit{II})}$ takes into consideration the phases accumulated during the time $T_N$ in both paths, and $b_{m_F}^{(j)}$ describes the populations of each mode after the rf pulse. Note that the phase $\xi^{(j)}$ has now been included in $\theta_{m_F}^{(\textit{I},\textit{II})}$ and $\varphi_{m_F}^{(\textit{I},\textit{II})}$. To observe the interference pattern, it is necessary that the final relative velocity between the two interfering wave packets is much smaller than the expansion velocity of each one of them, so that the interfering wave packets are well overlapped~\cite{13machluf2013coherent}. In our experiments, the relative velocity of wave packets is approximately \SI[per-mode=symbol]{0.4}{\micro\metre\per\milli\second}, and the expansion velocity is approximately \SI[per-mode=symbol]{3.0}{\micro\metre\per\milli\second}. Thus the requirement is totally fulfilled.

\subsubsection{\textit{\textbf{Observed interfering state $\left|\psi_{t}\right\rangle$ [Fig.~\ref{fig:procedure}(d)]}}}
Finally, after TOF, the final state $\left|\psi_{t}\right\rangle$ can be re-expressed as
\begin{align}
	\left|\psi_{t}\right\rangle &= \frac{1}{\sqrt{2}} \sum_{j=\textit{I},\textit{II}} e^{-i\varphi_{r}^{(j)}} \left\vert \psi^{(j)}\right\rangle \left\vert {p^{(j)},y^{(j)}}\right\rangle 
\end{align}
where $e^{-i\varphi_{r}^{(\textit{I}, \textit{II})}}$ are the overall phases of the two wave packets, and $\left\vert \psi^{(\textit{I}, \textit{II})}\right\rangle = \sum_{m_F} b_{m_F}^{(\textit{I}, \textit{II})} e^{-i\phi_{m_F}^{(\textit{I}, \textit{II})}} \left|m_F\right\rangle$ are the wave functions for the two superposition states, in which $\phi_{m_F}^{(\textit{I}, \textit{II})}= \theta_{m_F}^{(\textit{I},\textit{II})}+\varphi_{m_F}^{(\textit{I},\textit{II})}-\varphi_{r}^{(\textit{I}, \textit{II})}$. 

The interference pattern is described by the atomic density distribution at position $y$
\begin{align}\label{eq:interference0}
	\rho (y) =\left\vert\left\langle y \vert \psi_{t} \right\rangle\right\vert^2 = 1 + V_N(T_d,T_N) \cos(\kappa y+\eta)
\end{align}
where $\kappa=({p}^{(\textit{II})}-{p}^{(\textit{I})})/\hbar$ is the spatial frequency of the fringe, and $\eta=\varphi_{r}^{(\textit{II})}-\varphi_{r}^{(\textit{I})}$ is the overall phase of the interference fringe. $V_N(T_d,T_N)=\lvert\langle \psi^{(\textit{I})} \left\vert \psi^{(\textit{II})}\right\rangle\rvert$ is the visibility of the interference fringe, which is influenced by different relative phase evolution between the same internal modes in the two paths.

Here the relative phase is accumulated in two ways. First, an overall phase of each of the two paths (or the wave packets in the two paths) is obtained, which is mainly determined by the duration $T_d$ of the gradient magnetic field. This overall relative phase determines the profile of the interference pattern. Second, the relative phases between different states within one path (wave packet) emerge during $T_N$ and the TOF, which can be characterized by the state-dependent phase evolution rate $\omega_{m_F}^{(\textit{I}, \textit{II})}$.  We address our attention to the visibility. It shows a periodic dependence on $T_N$, which forms the so-called time domain interference. 
A remarkable feature of the multimode interferometer is the enhancement of resolution, which is defined as (fringe period)/(full width at half maximum), through the increase in internal modes. Note that so far we have not taken the shape of the wave packets into consideration and we give the formula for $V_N$. The wave-packet function and the dependence of visibility on $T_N$ and $T_d$ will be discussed in detail in Sec.~\ref{sec:time domain fringe}.

\section{Experimental demonstration of the interferometer}\label{sec:experiment}

We prepare a BEC of \textsuperscript{87}Rb with typically \num{1.0e5} atoms in the $\left|F=2, m_F=2\right\rangle$ state as our initial state, which is confined in a magnetic and optical hybrid harmonic trap with the trap frequencies $(\omega_x, \omega_y, \omega_z)=2\pi\times (\SI{28}{\hertz}, \SI{55}{\hertz}, \SI{65}{\hertz})$. The preparation of the BEC is similar to our previous experiments\cite{12.2linixao,12.3Jin_2019}. The magnetic sublevels are considered as modes in the interferometer, with different phase evolution rates in the magnetic field. Therefore, the hyperfine state number $F$ has determined our maximum number of modes $N_{\text{max}}=5$ ($m_F=-2,\dots,2$). Our initial BEC temperature $\simeq \SI{50}{\nano\kelvin}$ guarantees the coherence of the atomic wave packets. The atoms experience a magnetic field with amplitude of $B_0$ = \SI{443}{\milli\gauss}, which resulted in a Zeeman split of $\simeq \SI{310}{\kilo\hertz}$, and gradient of $\partial_y B=\SI[per-mode=symbol]{12.4}{\gauss\per\centi\metre}$. 

In the double-path operation, we choose magnetic sublevels $\left|F=2, m_F=1,2\right\rangle$ as the states $\left|\textit{I},\textit{II}\right\rangle$, which can be achieved by a nonadiabatic Majorana transition. Here the fine tunable spin projection of the nonadiabatic Majorana transition is achieved by precisely modulating the rotation of the magnetic field, where the transition time is chosen as \SI{15.0}{\micro\second}\cite{13tang2019implementation}. It transfers the initial state into ${|m_F=2\rangle}$ and ${|m_F=1\rangle}$ almost equally, leaving other states negligible, which is better than the rf as a one-to-two beam splitter\cite{13tang2019implementation}. After the Majorana transition and evolving for time $T_d$, the two wave packets are spatially separated only for tens of nanometers, which is approximately 1\% of the BEC size. In the multimode operation, the rf pulse behaves more effectively to divide both the two wave packets ($\left|\textit{I}\right\rangle$ and $\left|\textit{II}\right\rangle$) into five states as evenly as possible\cite{7.1PhysRevATPJ}. So we choose rf in the multimode operation and set the duration of the rf pulse as $\tau_R=\SI{10.0}{\micro\second}$, the frequency of which is set to the resonant frequency \SI{310}{\kilo\hertz} to match the Zeeman split. The pulse amplitude is a selected constant in order to achieve a one-to-five beam splitter, which makes both $\left|\textit{I}\right\rangle$ and $\left|\textit{II}\right\rangle$ transfer into all five Zeeman states. After the rf pulse, the populations of the five modes from $m_F=2$ to $-2$ are 14, 19, 27, 24, and 16\%, respectively, which is even enough for the following measurements.

\begin{figure}
	\includegraphics[width=\linewidth]{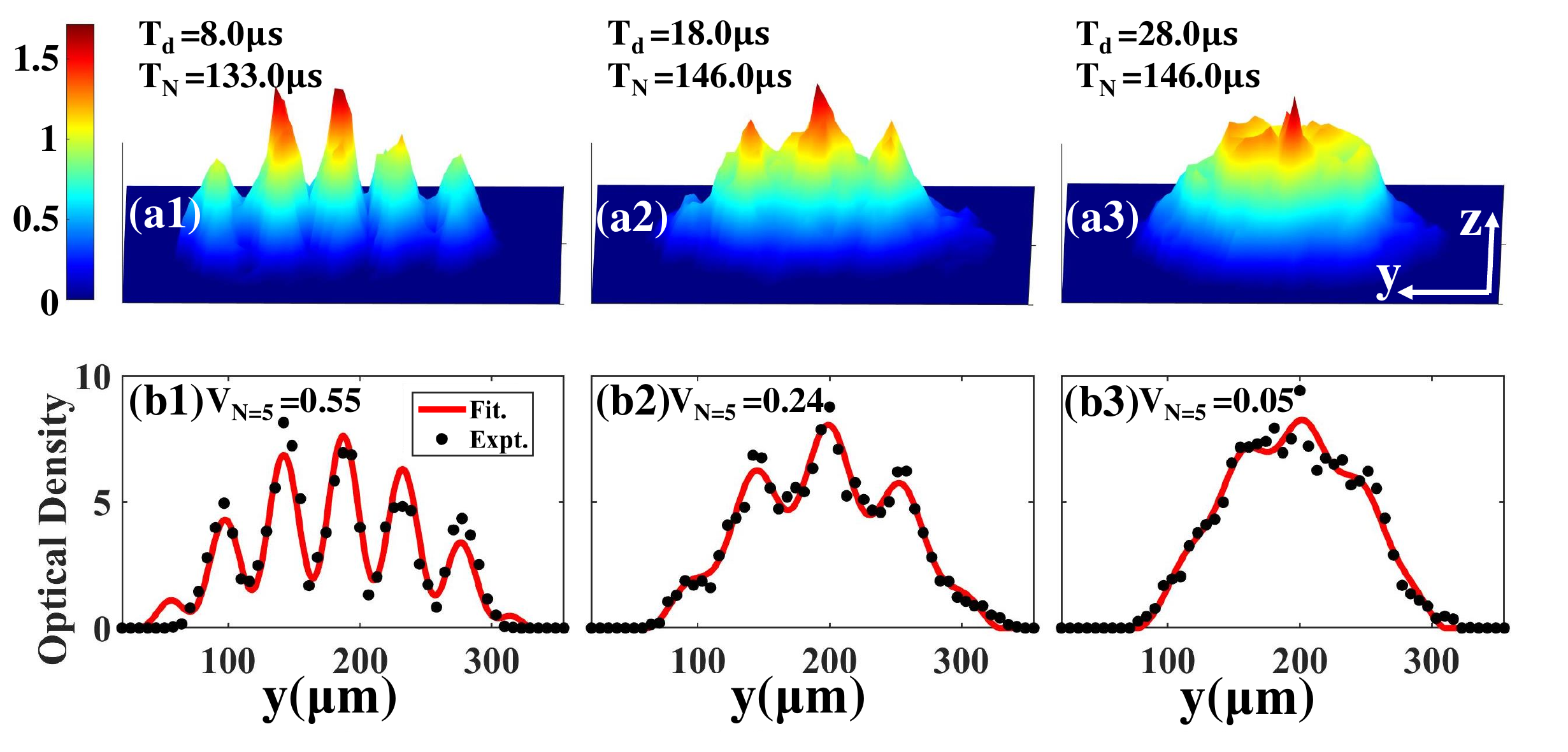}
	\caption{
		(Color online) (a1)-(a3) Single-shot spatial interference pattern with $N=5$ interference modes after $T_{\textit{TOF}}=\SI{26}{\milli\second}$, in which (a1) $T_d=\SI{8.0}{\micro\second} ,T_N=\SI[]{133.0}{\micro\second}$, (a2) $T_d=\SI{18.0}{\micro\second},T_N=\SI[]{146.0}{\micro\second}$ and (a3) $T_d=\SI{28.0}{\micro\second},T_N=\SI[]{146.0}{\micro\second}$, with different relative phase between adjacent modes. (b1)-(b3) Data (black points) are got by integrating the image in panels (a1)–(a3) along the  $z$ direction and they are fitted by Eq.~\eqref{eqn:interference} (red solid lines). The visibilities are 0.55, 0.24, and 0.05, respectively.}
	\label{fig:interference}
\end{figure}	

The observed fringe is actually a superposition of the interference fringes of different modes. And the visibility depends on the relative phase between the interference fringes of different modes $\left\vert m_F\right\rangle$, where the fringe of each mode is formed by the same mode in the two paths. The relative phase can be investigated by tuning $T_d$ and $T_N$. When all the modes are totally in phase, the visibility could reach $V_N=1.0$ in theory. Setting $T_d=\SI{8.0}{\micro\second}$ and $T_N=\SI{133.0}{\micro\second}$, we observed the clearest interference, as shown in Fig.~\ref{fig:interference}(a1). If the modes are partially in phase, the visibility drops but remains a nonzero value [Fig.~\ref{fig:interference}(a2)], in which $T_d=\SI{18.0}{\micro\second}$ and $T_N=\SI{146.0}{\micro\second}$. And when all the fringes are complementary in space the pattern without any fringes is observed as shown in Fig.~\ref{fig:interference}(a3), where $T_d=\SI{28.0}{\micro\second}$ and $T_N=\SI{146.0}{\micro\second}$. All of the images in this paper are taken by absorption imaging after \SI[]{26}{\ms} TOF.

To take into account the shape of the atomic cloud and the imperfection in the overlap between wave packets of the two paths, we model the atomic clouds as the sum of two Thomas-Fermi wave packets\cite{13machluf2013coherent}
\begin{align}\label{eqn:interference}
	\Lambda= \rho\sum_{j=\textit{I},\textit{II}} F^{(j)}(y)
\end{align}
Here {\small $F^{(j)}(y) = A^{(j)} \text{max}\left\{ \left[ 1-(y-y_0^{(j)})^{2}/2(\sigma^{(j)})^2 \right]^{{3}/{2}}, 0 \right\}$} is the Thomas-Fermi function, $A^{(j)}$ is the amplitude, $y_0^{(j)}$ represents the center of the atomic cloud, and $\sigma^{(j)}$ is the width of the wave packet. We integrate the data in Fig.~\ref{fig:interference}(a1) along the $z$ axis and fit it by Eq.~\eqref{eqn:interference}, which shows a higher visibility $V_{N=5}=0.55$ with a fringe spacing of $\lambda=2\pi/\kappa=\SI{46}{\micro\metre}$, as shown in Fig.~\ref{fig:interference}(b1). However, the visibility reduces to $V_{N=5}=0.24$ [Fig.~\ref{fig:interference}(b2)] and $V_{N=5}=0.05$ [Fig.~\ref{fig:interference}(b3)] with other values of $T_d$ and $T_N$. This means the relative phase can influence the visibility and we will discuss this in the following sections.

\section{The resolution of time domain interference}\label{sec:time domain fringe}

\subsection{Formulating the time domain interference}\label{sec:formulatingVisibility}

The most important difference between a double-path multimode interferometer and a traditional double-path interferometer lies in the fact that the visibility is modulated by the phase difference between two paths, and thus the visibility exhibits a periodic change and causes the emergence of the so-called time domain interference. The peak widths in time domain fringe are determined by the number of modes $N$ as well as the initial relative phase between the same mode in two paths $\Delta\theta_{m_F}$. Now in this section, following Eq.~\eqref{eq:interference0} in Sec.~\ref{subsec:InterferenceAnalysis}, we derive the formula for the visibility $V_N(T_d,T_N)$ and discuss its dependence on $T_d$ and $T_N$.

The visibility of the spatial interference fringe is\cite{schmiedmayer1994magnetic}
\begin{align}\label{eqn:V} 
	V_N(T_d,T_N) \nonumber
	&= \left\vert\left\langle \psi^{(\textit{I})} \left\vert \psi^{(\textit{II})}\right.\right\rangle\right\vert \\ \nonumber
	&= \left|\sum_{m_F=3-N}\limits^{2}{b_{m_{F}}^{(\textit{II})}}{b_{m_{F}}^{(\textit{I})}} e^{-i[\phi_{m_F}^{(\textit{I})}-\phi_{m_F}^{(\textit{II})}]}\right| \\ 
	&= \left|\sum_{m_F=3-N}\limits^{2}{b_{m_{F}}^{(\textit{II})}}{b_{m_{F}}^{(\textit{I})}} e^{-i\Delta\phi_{m_F}}\right|	
\end{align}
where $N=3,4,5$ is the number of modes. When considering the phase evolution, only the relative phase is relevant. Therefore, we use the $\left|m_F=2\right\rangle $ state as a reference and denote the phase of each component as $\phi_{m_F}^{(j)}=\varphi_{2\leftarrow m_F}^{(j)}+\theta_{m_F}^{(j)}$ and $\varphi_{2\leftarrow m_F}^{(j)} = \varphi_{m_F}^{(j)}-\varphi_{2}^{(j)}= (2-m_F)\omega^{(j)} T_N$, where $\omega^{(j)}$ is the relative phase evolution rate between adjacent modes in path $j$. The relative phase evolution rate between the two paths is defined as $\Delta \omega=\omega^{(\textit{I})}-\omega^{(\textit{II})}$, where $\omega^{(\textit{I})}$ and $\omega^{(\textit{II})}$ are the phase evolution rates in each path, respectively. The relative phase of the $\left\vert m_F\right\rangle$ states in two paths is $\Delta\phi_{m_F} =\phi_{ {m_F}}^{(\textit{I})}-\phi_{ {m_F}}^{(\textit{II})}=(2-{m_F})\Delta \omega T_N + \Delta\theta_{m_F}$, where the relative phase evolution rate $\Delta\omega$ is dependent on $T_d$, and $\Delta\theta_{m_F} = \theta_{m_F}^{(\textit{I})}- \theta_{m_F}^{(\textit{II})}$ is the initial relative phase of the $\left\vert m_F\right\rangle$ modes in the two superposition states. The relative phase between two adjacent components is
\begin{align}\label{eqn:phi}
	\begin{split}
		\Delta\phi
		&=(\Delta\phi_{m_F}-\Delta\theta_{m_F})-(\Delta\phi_{m_F-1}-\Delta\theta_{m_F-1})\\ 
		&=\Delta \omega T_N
	\end{split} 
\end{align}

From Eq.~\eqref{eqn:phi} we know that the relative phase $\Delta\phi$ (and thus the visibility $V_N$) is periodically modulated with the period $2\pi/\Delta\omega$, leading to the time domain fringe. So far, we can see that the time domain fringe is influenced by the number of modes, the initial relative phase, and the relative phase evolution rate between two paths  $\Delta\omega$, which will be discussed with experiments in the following subsections.

\begin{figure}
	\includegraphics[width=\linewidth]{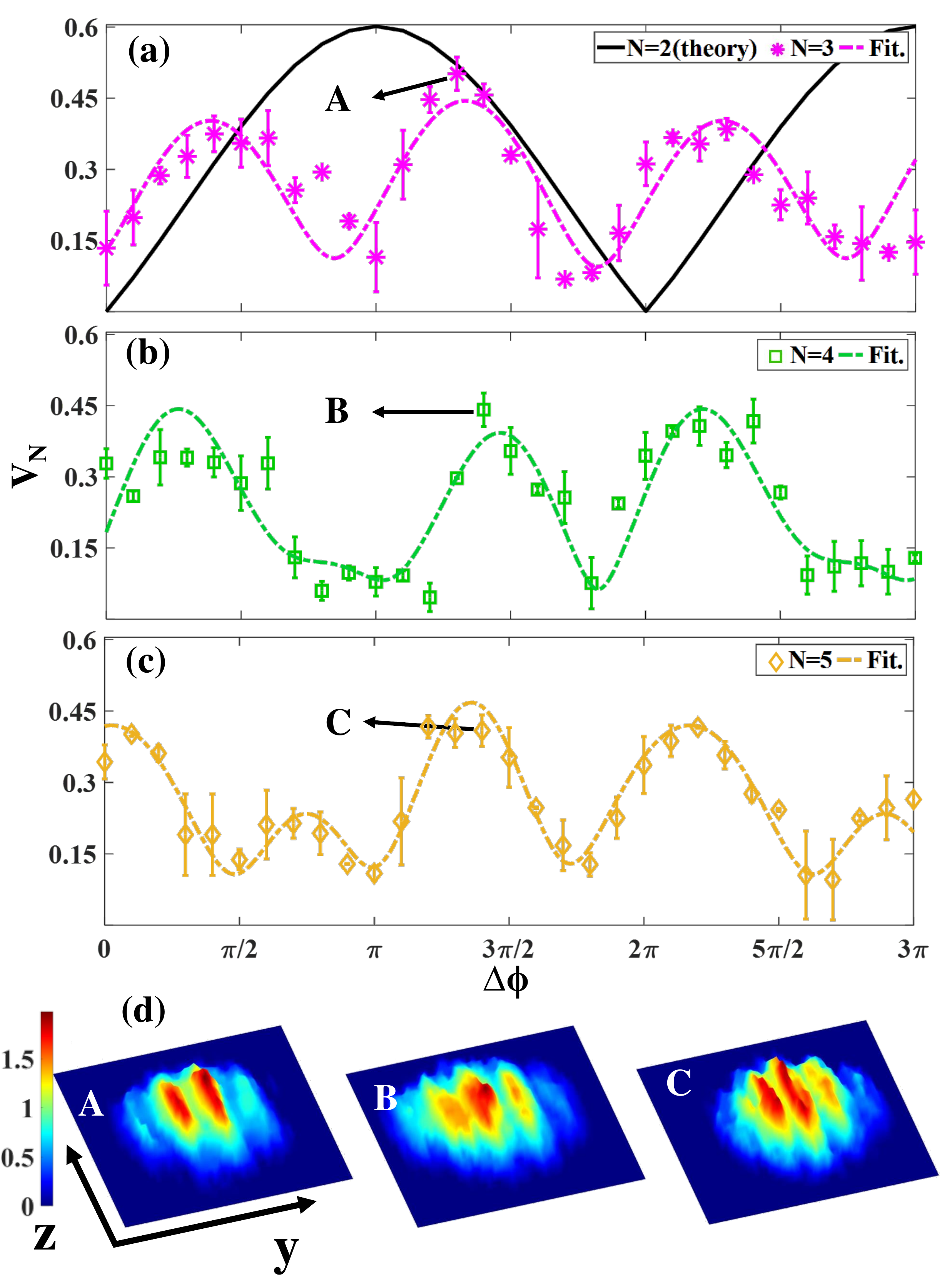}
	\caption{(Color online) The oscillation of visibility $V_N$ with $\Delta\phi$ with (a) $N=3$ modes, (b) $N=4$ modes and (c) $N=5$ modes, with $T_d = \SI[]{42.5}{\micro\second}$. The points and the dashed lines show the experimental data and fit respectively. Each data point is a statistical average of four repeated experiments and the error bar shows the standard deviation. In the fitting, free parameters are ${b_{m_{F}}^{(\textit{II})}}$, ${b_{m_{F}}^{(\textit{I})}}$ and $\Delta\phi_{m_F}$. The theoretical result of a two-mode situation is shown by the black solid line in panel (a) for comparison. We set the phase to be zero at $T_N = \SI[]{110.0}{\micro\second}$ as a reference and the data are measured from $T_N = \SI[]{110.0}{}$ to $ \SI[]{140.0}{\micro\second}$. $\Delta\omega = \SI[per-mode=symbol]{0.31\pm 0.03}{\radian\per\micro\second}$ is used in all three measurements. (d)The TOF image near the main peak with different number of modes $N$. The images in A, B, and C correspond to points A, B, and C in (a), (b), and (c), respectively.}\label{fig:beta}
\end{figure}

\subsection{Sharper time domain fringes with more modes}\label{subsec:N-effect}

We experimentally study how the structure of the time domain fringes changes with the number of modes $N$, as shown in Fig.~\ref{fig:beta}. Different modes $N$ are achieved by changing the power of the rf, while the rf duration is unchanged. $T_d$ is chosen as \SI{42.5}{\micro\second} to maximize the visibility (see Appendix.~\ref{sec:Appendix0}). In Fig.~\ref{fig:beta}(a), compared to the time domain fringe of two-path two-mode interference (solid black line), the time domain fringe of the double-path three-mode interference exhibits two peaks in one period. According to Figs.~\ref{fig:beta}(b) and \ref{fig:beta}(c), a more complex internal structure of time domain fringes emerges, in which more peaks appear in each cycle and the fringes become sharper. This comes from two mechanisms: high-order harmonics induced by multimodes and the influence of a different initial relative phase, which will be discussed in detail in the next subsection.

By fitting the data with Eq.~\eqref{eqn:V}, we find the periods of time domain fringe for different $N$ are almost the same, and $\Delta\omega$ is measured as \SI[per-mode=symbol]{0.31\pm 0.03}{\radian\per\micro\second} in Fig.~\ref{fig:beta}, which nearly matches the theoretical estimation $\Delta\omega_{\textit{est}} = \SI[per-mode=symbol]{0.27 \pm 0.03}{\radian\per\micro\second}$. Meanwhile, as the number of modes increases from Fig.~\ref{fig:beta}(a) to \ref{fig:beta}(b), the visibility of the main peak (A, B, C) remains essentially unchanged around $0.45$, where the corresponding TOF images are shown in Fig.~\ref{fig:beta}(d), respectively. We get the full width at half maximum of the main peaks and then the corresponding resolutions are measured as (a) \num{2.4 \pm 0.3}, (b) \num{2.7 \pm 0.3}, and (c) \num{3.0 \pm 0.3}. They are all larger than the resolution of an ideal double-path two-mode interferometer \num{1.5}. The deviation of the visibility from the ideal value is due to the contribution of the residual magnetic-field fluctuations, which is estimated as \SI{1}{\milli\gauss}, and time control accuracy (see Appendix.~\ref{sec:Appendix2}). So with Fig.~\ref{fig:beta}, we demonstrate that as the number of modes increases, the main peak visibility of the fringes of the time domain interference is basically unchanged, and the phase resolution is gradually improved.

\subsection{Higher resolution with appropriate initial phase}\label{subsec:initialPhase-effect}

It is necessary to point out that the initial phase of each mode could influence the shape of time domain fringes.
In order to give a general discussion, we do not confine ourselves to the five-mode situation and we consider an arbitrary number of modes. The wave function of a multimode superposition state can be defined as\cite{12margalit2015self}
\begin{align}
	\left|\psi^{(j)}\right\rangle=\zeta(y)\sum_{l=1}\limits^{N}e^{-i{({\omega_l}^{(j)}}T_N+{\theta_l}^{(j)})}\left\vert l\right\rangle
\end{align}
where $j=\textit{I},\textit{II}$ represents the two paths, $\zeta(y)=\frac {\chi(y)}{\sqrt{2N}} e^{-ik^{(j)}y}$ represents a moving wave packet, ${\chi(y)}$ is a (one-dimensional) localized normalized wave function, and $k^{(j)}$ is the wave vector of the $j$-th wave packet. Here, the notion $l$ corresponds to the magnetic sublevels $\left|m_F\right\rangle$ used in previous discussions. ${\omega_l}^{(j)}=l\omega^{(j)}$ represents the phase evolution rate of the $l$-th mode ($\left\vert l\right\rangle$ state) in path $j$. When the two wave packets are fully overlapped, the density distribution is given by $\left|\left|\psi^{(\textit{I})}\right\rangle + \left|\psi^{(\textit{II})}\right\rangle\right|^2$, leading to the visibility
\begin{align}\label{eqn:V-n}
	V_N = \left|\sum_{l=1}\limits^{N}{\frac {1}{N}} e^{-i(l\Delta\phi+\Delta\theta_l)}\right|
\end{align}
where $\Delta \phi=\Delta \omega T_N$ is the relative phase of the fringes between adjacent modes and $\Delta \omega$ is the relative phase evolution rate. The $\Delta\theta_l={\theta_l}^{(\textit{I})}-{\theta_l}^{(\textit{II})}$ is the initial relative phase of the $\left\vert l\right\rangle$ state in two wave packets.  

\begin{figure}
	\includegraphics[width=\linewidth]{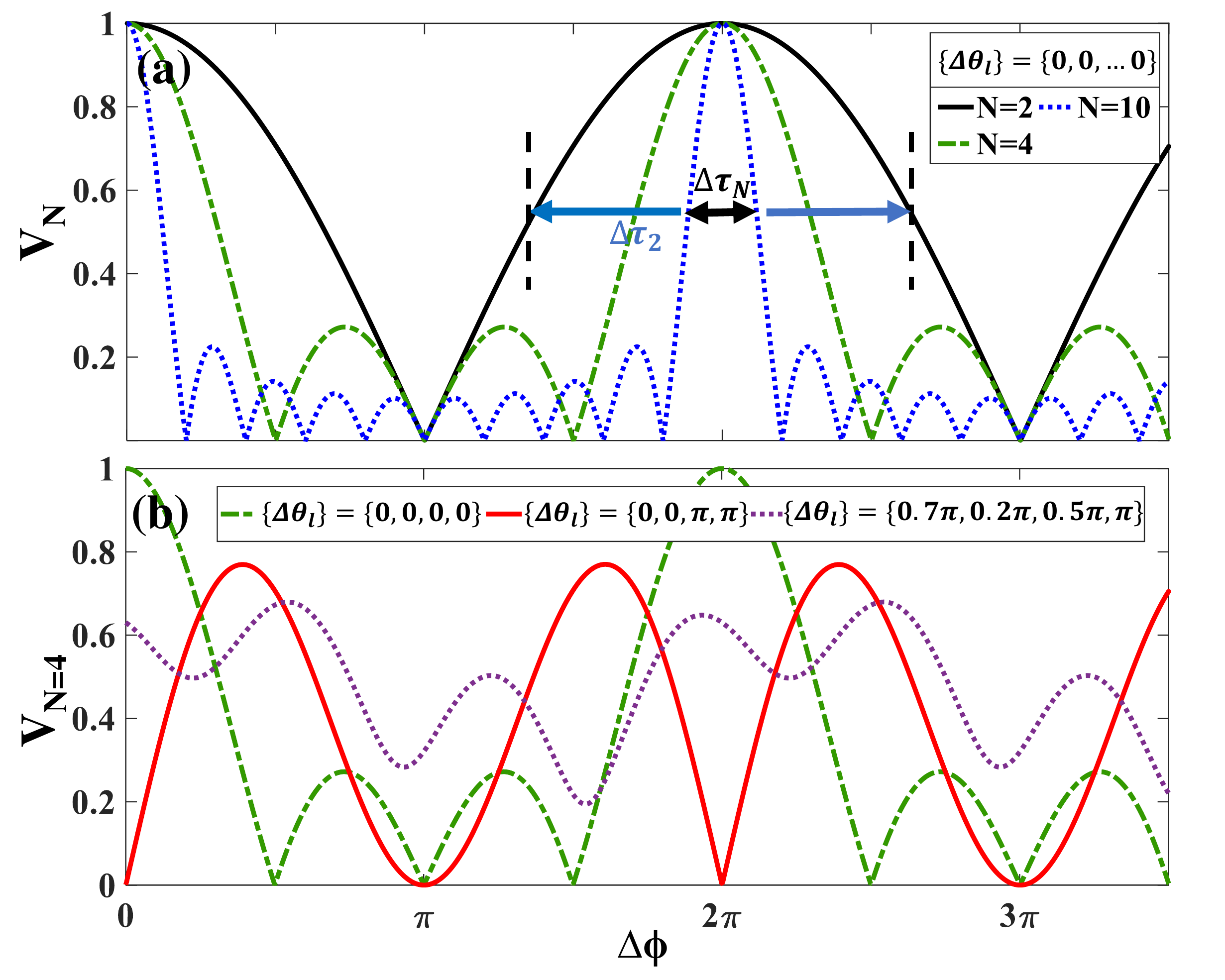}
	\caption{(Color online) The dependence of visibility on the number of modes $N$ and initial relative phase of the same mode in two paths $\Delta\theta_l$. (a) Dependence on $N$ in a situation that the relative initial phases $\Delta\theta_l$ are all zero. The full width at half maximum $\Delta\tau_N$ of the $N$-mode fringe is $2/N$ times that of the two-mode fringe. (b) Dependence on $\Delta\theta_l$ using $N=4$ as an example. The green dashed line, red solid line, and purple dotted line show the fringes with $\left\{\Delta\theta_l\right\} = \left\{\Delta\theta_1, \Delta\theta_2, \Delta\theta_3, \Delta\theta_4 \right\} = \left\{0, 0, 0, 0\right\}$, $\left\{\Delta\theta_l\right\}  = \left\{0,0,\pi,\pi\right\}$, and $\left\{\Delta\theta_l\right\}  = \left\{0.7\pi,0.2\pi,0.5\pi,\pi\right\}$, respectively.}\label{fig:visibility}
\end{figure}

The sharpness of the peaks (and thus the resolution) varies for different $\Delta\theta_l$. Fig.~\ref{fig:visibility}(a) is under the condition that the phases $\Delta\theta_l$ are all the same for any $l$. In that case, if we denote $\Delta\omega T_N=2n\pi/N$, then when $n$ is a non-negative integer but is not the multiple of $N$, the visibility $V_N=0$, while the visibility achieves $V_N=1$ when $n$ is the multiple of $N$ and a major peak is observed in this case. As a result, the peak of the visibility gets much sharper as $N$ increases, which leads to a higher resolution. It is the harmonics that cause the peak width to decrease with the number of modes increasing in this case \cite{3.4Pikovski2017clock}. A remarkable feature of our interferometer is that the enhancement of resolution with $N/2$ times is achieved without reduction in visibility.

However, if the initial relative phase of each state $\Delta\theta_l$ varies from mode to mode [Fig.~\ref{fig:visibility}(b) serves as an example at $N=4$], the visibility could reach neither the maximum $V_N=1.0$ nor the minimum $V_N=0$. Meanwhile, the time domain fringe shows more than one main peak in one period. For some specific configurations in which an initial relative phase is chosen, for example, $\left\{\Delta\theta_l\right\} = \left\{\Delta\theta_1, \Delta\theta_2, \Delta\theta_3, \Delta\theta_4\right\} = \left\{0, 0, \pi, \pi\right\}$, the interference fringes with two main peaks in each period are observed, as shown by the red solid line in Fig.~\ref{fig:visibility}(b). Therefore, the resolution deteriorates more or less, which explains what we observed in Fig.~\ref{fig:beta}. Meanwhile, if the initial relative phase is chosen randomly as, e.g., $\left\{\Delta\theta_l\right\} = \left\{0.7\pi,0.2\pi,0.5\pi,\pi\right\}$, there are no interference fringes with obvious periodic characteristics. Therefore, the initial phase $\Delta\theta_l$ needs to be well controlled to achieve the highest possible visibility and clear interference fringe in the time domain. 

\subsection{Investigating relative phase evolution rate $\Delta\omega$}\label{subsec:omega-effect}

\begin{figure}
	\includegraphics[width=\linewidth]{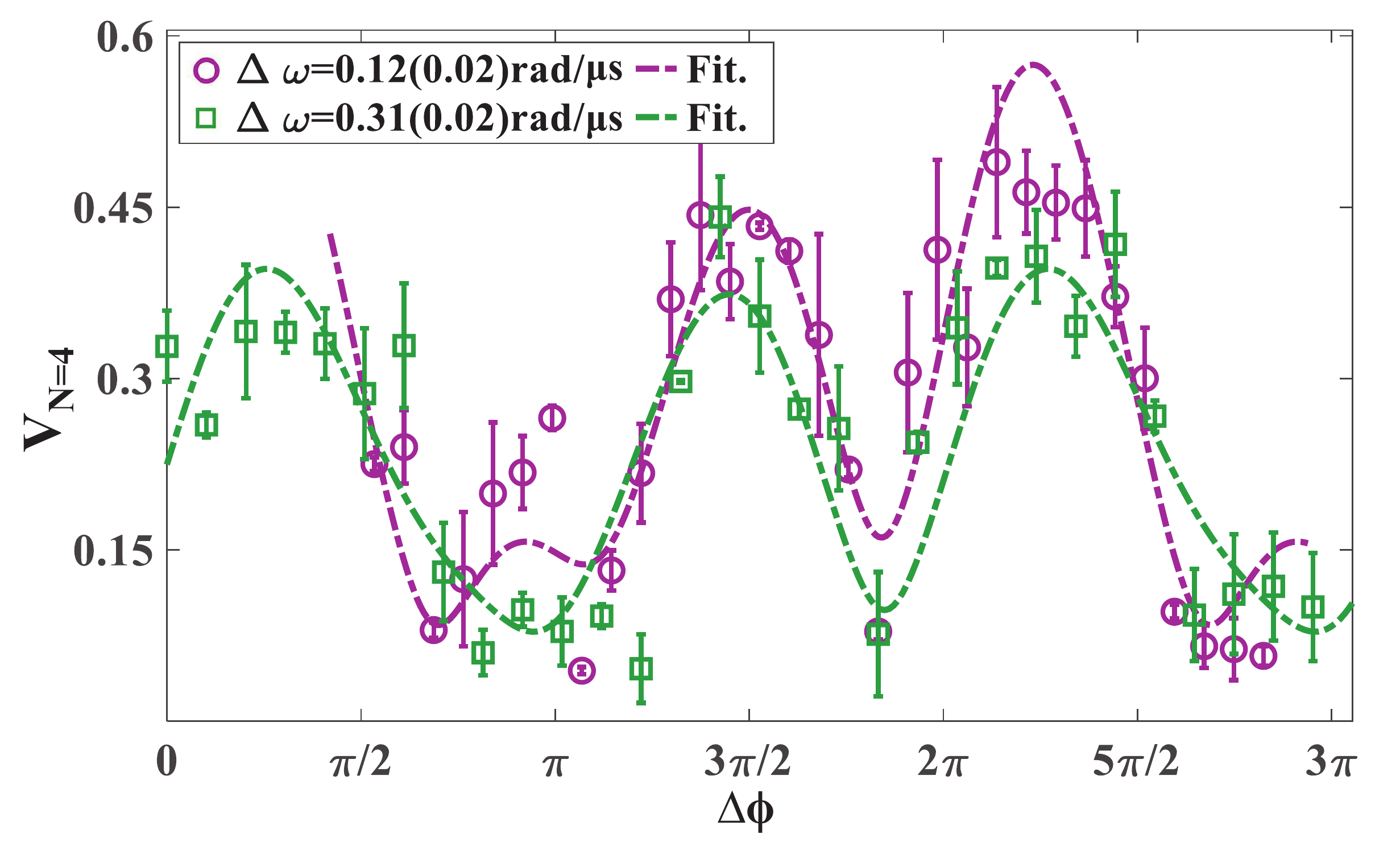}
	\caption{
		(Color online) The oscillation of visibility with $\Delta\phi$ under the conditions of different $\Delta\omega$ at $N=4$. Dashed lines show the fit results. The green (purple) points represent $T_d = \SI[]{42.5}{\micro\second}$ ($T_d = \SI[]{17.0}{\micro\second}$) and corresponding $\Delta\omega=\SI[per-mode=symbol]{0.31\pm 0.02}{\radian\per\micro\second}$ ($\Delta\omega=\SI[per-mode=symbol]{0.12\pm 0.02}{\radian\per\micro\second}$). Each data point is a statistical average of four repeated experiments and the error bar shows the standard deviation. We set the phase of the green line at $T_N = \SI[]{110.0}{\micro\second}$ to be zero as a reference, and the data are measured from \SI[]{110.0}{} to \SI[]{140.0}{\micro\second}. For purple points we set the valley phase to be the same as the valley value of the green line around $2\pi$ and the data are measured from \SI[]{100.0}{} to \SI[]{166.0}{\micro\second}.
	}
	\label{fig:omega}
\end{figure}

When one studies the traditional double-path interference, the difference of the phase evolution rates between the two paths results in a change in the period of the spatial interference fringe. However, when considering a double-path multimode interference in the time domain we address our attention to how the difference between two paths influences the period of the time domain fringe, which corresponds to $2\pi/\Delta\omega$. $\Delta\omega$ is mainly affected by $T_d$, i.e., the double-path operation, since the path difference lies in the difference between the momenta of the two wave packets and their positions in the harmonic trap.

To show the effects of $\Delta\omega$, we measured the time domain fringes for different $T_d$ for number of modes $N=4$, as shown in Fig.~\ref{fig:omega}. The purple and green lines are obtained with $\Delta\omega=\SI[per-mode=symbol]{0.12\pm0.02}{\radian\per\micro\second}$ ($T_d=\SI{17.0}{\micro\second}$) and with $\Delta\omega = \SI[per-mode=symbol]{0.31 \pm 0.02}{\radian\per\micro\second}$ ($T_d=\SI{42.5}{\micro\second}$), respectively. The resolution is measured as \SI{2.7\pm 0.3} and, the purple and green lines show a similar period with respect to phase $\Delta\omega T_N$, so the period $2\pi/\Delta\omega$ of the green line in the time domain is lower than the half of the purple line with respect to time $T_N$. As a result, the data shown with purple lines give a more accurate measurement of the phase $\Delta\phi$, as any small errors in the duration of $T_N$ are less severe for the slower dynamics. In general, when increasing the difference of the two paths in the double-path multimode interferometer by increasing $T_d$, we can decrease the period of the time interference fringe $2\pi/\Delta\omega$ and improve the measurement of $\Delta\phi$.

\section{Coherence analysis}\label{sec:discussion}

\begin{figure}
	\includegraphics[width=\linewidth]{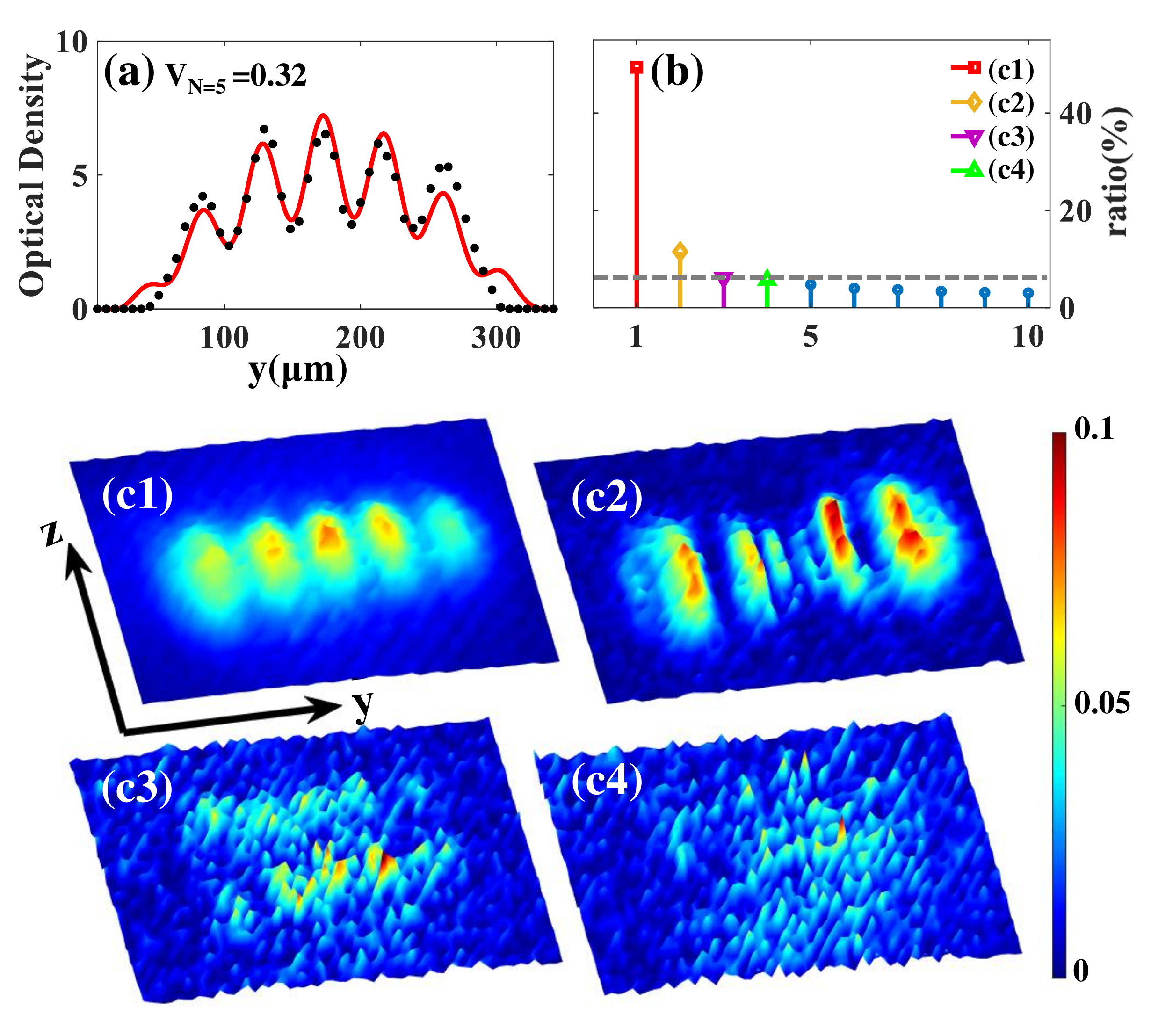}
	\caption{
		(Color online) (a) Averaged interference fringes of 42 consecutive measurements. The black points are the experimental data integrated along the $z$ direction, and the red line is the fitted curve by Eq.~\eqref{eqn:interference}, and we get the visibility 0.32. (b) The percentage ratio of the first ten principle components (PCs). The gray dashed line shows the level of background noise ($\simeq 5\%$) and indicates the threshold for distinguishing important PCs. PCs of interest are highlighted with different colors, which correspond to panels (c1)-(c4) from left to right. Note that the first four PCs are significantly reduced as the order of PC increases. (c1) Average of interference fringes. (c2) and (c3) Spatial fluctuations along $y$ and $z$axes, respectively. They together represent atom cloud position fluctuations in the $y$-$z$ plane. (c4) Noise sources from imaging light.
	}
	\label{fig:interference0}
\end{figure}

In this section, we analyze the coherence of the fringes and the robustness of the measurements and, in particular, whether the fringes move and whether the visibility changes from shot to shot, in order to support that the interference is robust. We take the situation shown in Fig.~\ref{fig:interference}(a1) as an example, where 42 consecutive experiment data sets were taken under identical conditions. The visibility of the average reduces to $V_{N=5}=0.32$ as shown in Fig.~\ref{fig:interference0}(a) from the single-shot visibility $V_{N=5}=0.55$ as shown in Fig.~\ref{fig:interference}(a). The fact that we can still observe clear fringes in the averaged image indicates that the relative phase $\Delta\phi$ is quite stable, and we attribute the decrease of visibility from single shot to average to the noises in experiments. 

In order to understand the contribution of different noise sources, following our previous work in Refs.~\cite{Cao:19,niu2018optimized}, we perform a principle component (PC) analysis on the data sets. Fig.~\ref{fig:interference0}(b) shows percentages of the total variance associated with the first ten PCs. We find that the components of higher than the fourth order are negligible because they are below the level of background noise $\simeq 5\%$. We reconstruct the contribution of the first four PCs, which are shown in Figs.~\ref{fig:interference0}(c1)-\ref{fig:interference0}(c4). Fig.~\ref{fig:interference0}(c1) exhibits the main features of the original experimental image. The second and the third PCs shown in Figs.~\ref{fig:interference0}(c2)-\ref{fig:interference0}(c3), the percentages of which are of the same order of magnitude, reflect the position uncertainty of the fringes along the $y$ axis and $z$ axis, respectively. They together represent the position fluctuations of the atomic clouds position in the $y$-$z$ plane. The noises make the position fluctuations slightly anisotropic due to different eigenvalues. They come from the system noises such as the electrical noises of the rf coupling circuit and magnetic-field control circuit, as well as the fluctuations of the gradient magnetic field $\partial_y B$. The fluctuations in the atomic region of Fig.~\ref{fig:interference0}(c4) are comparable to the background noise, which could be attributed to the photon shot noise and CCD camera dark current noise.

\section{Conclusion}\label{sec:conclusion}

We demonstrated the principles of a double-path multimode matter wave interferometer, which is different from traditional double-path interference with a single component in each path. One can observe not only spatial interference fringes, but also time domain interference of the visibility. Meanwhile, we experimentally demonstrated its performance with the number of modes being 3, 4 and 5. The period of time domain fringes is independent of the number of modes. However, a remarkable feature of the double-path multimode interferometer is the enhancement of phase measurement resolution in the time domain with the increasing of the number of modes, where the double-path five-mode interferometer is enhanced twice compared to an ideal double-path two-mode interferometer [Fig.~\ref{fig:beta}(c)]. And compared with a traditional five-path single-mode interferometer our double-path five-mode interferometer does not reduce the atom number for effective interference (the atom number for traditional readout with the Stern-Gerlach process is reduced to $1/N$ for $N$ modes) and thus suppresses the shot noise. Last but not least, the relative phase evolution rate $\Delta \omega$ can be controlled by adjusting the difference between the two paths accumulated in the $T_d$ stage. Furthermore, the interferometer is implemented in a trap with a weak magnetic field, making it easy to be integrated into a miniaturized interferometer on an atomic chip and convenient to be combined with an optical lattice during the process of the phase evolution.

\begin{acknowledgements}
	We thank Weidong Li, Baoguo Yang, Peng Zhang, Xinhao Zou, and Xiaopeng Li for helpful discussion. This work is supported by the National Basic Research Program of China ((Grant No. 2016YFA0301501) and the National Natural Science Foundation of China (Grants
	No. 61727819, No. 11934002, No. 91736208, and No. 11920101004).
\end{acknowledgements}

\appendix
\addcontentsline{toc}{section}{Appendices}\markboth{APPENDICES}{}
\begin{subappendices}
	
	\section{Choosing appropriate $T_d$ for clear time domain fringe}\label{sec:Appendix0}
	
	In the main text, we mainly discussed how the visibility is affected by $T_N$. In fact, $T_d$ also has an impact on visibility. The role of the $T_d$ is mainly reflected in two aspects: modifying initial phase $\Delta\theta_{m_F}$ and affecting the quality of the overlapping between two wave packets. By choosing appropriate $T_d$, we can get a time domain fringe with high maximum visibility.
	
\begin{figure}
		\includegraphics[width=\linewidth]{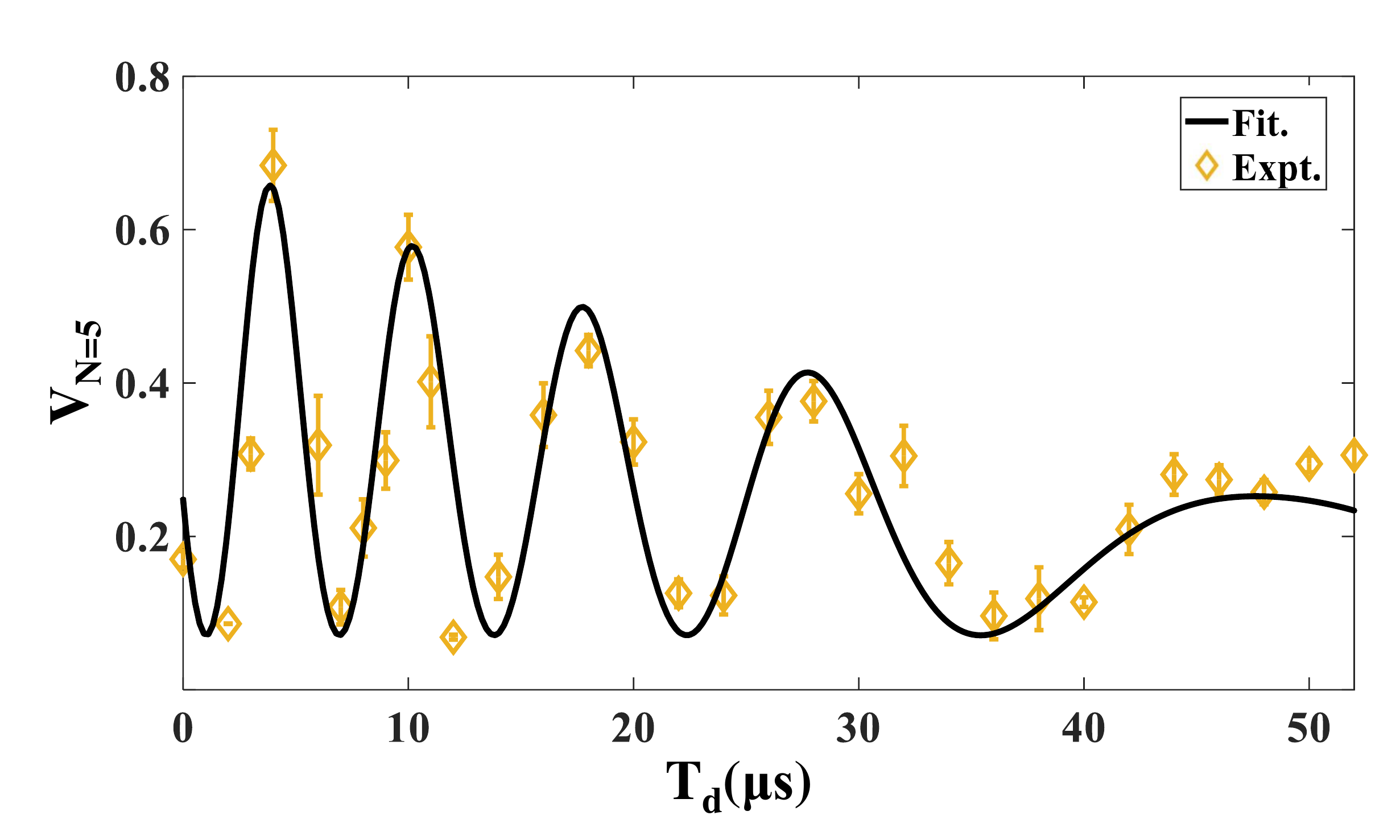}
		\caption{(Color online) The visibility $V_N$ vs $T_d$ when $T_N=\SI[]{146.0}{\micro\second}$ and $N=5$. The yellow points are the experimental data with average of three shots, and the black solid line is obtained by fitting the curve using Eq.~\eqref{eqn:fitt2}. }
		\label{fig:interference1}
	\end{figure}

	In order to demonstrate the effect of $T_d$, we keep $T_N=\SI{146.0}{\micro\second}$ unchanged, and measure the dependence of visibility on $T_d$, as shown in Fig.~\ref{fig:interference1}. It shows a clear trend that the visibility experiences a damping oscillation as $T_d$ increases. To account for this phenomenon, we model this process using the following empirical formula, Eq.~\eqref{eqn:fitt2}, to fit the experimental data.
	\begin{align}\label{eqn:fitt2}
		V_N=R\exp{(-\gamma{T_d})}[1+\cos(\omega_d-\varepsilon{T_d}){T_d}+\alpha]+\beta
	\end{align} 
	where $R$ is the amplitude and $\gamma$ is the damping factor. The phase evolution rate is modeled in a gradually changing way $\omega_d-\varepsilon T_d$, where $\varepsilon T_d$ represents a linear change. $\alpha$ represents the initial phase of the time domain fringe and $\beta$ is a bias. The oscillation amplitude gradually reduces, because the wave-packet overlap of the two paths becomes smaller.
	
	When a good set of initial phases $\Delta\theta_{m_F}$ is chosen, the interference fringe could reach a high maximum visibility, as demonstrated in Fig.~\ref{fig:visibility}(b). Therefore, we could obtain the appropriate set of $\Delta\theta_{m_F}$ by selecting $T_d$ at the peak position in the curve in Fig.~\ref{fig:interference1} to achieve a clear time domain interference fringe, which shows a high maximum visibility and a simple period. Considering that the frequency $\Delta\omega$ will be too small to measure in experiment if $T_d$ is too small, the smallest value for $T_d$ is chosen to be \SI{17.0}{\micro\second} as in Fig.~\ref{fig:omega}. On the other hand, to guarantee that two wave packets are sufficiently overlapped for high visibility, $T_d$ should be chosen to be less than \SI{50.0}{\micro\second}. As a result, the largest value for $T_d$ is chosen as \SI{42.5}{\micro\second} as in Fig.~\ref{fig:beta}.
	
	\section{Estimating $\Delta\omega$ from experiments}\label{sec:Appendix2}
	
	In order to estimate the relative phase evolution rate $\Delta \omega$ between two paths, we analyze the phase evolution of each Zeeman sublevel based on experimental parameters such as $T_d$ and $T_N$. After the double-path evolution $T_d$ and the multimode evolution $T_N$ in the harmonic trap and  the TOF with the expansion time $T_{\textit{TOF}}$, the phase of each state $\left|m_F \right\rangle$ in path $j=I,II$ can be expressed as $\varphi_{m_F}^{(j)}  = ({m}/{{2\hbar T_{\textit{TOF}}}}) {y_{m_F}^{(j)}}^2 + {m{\upsilon_{m_F}^{(j)}} }{y_{m_F}^{(j)}}/\hbar$ \cite{7fort2001spatial,6Simsarian2000phase,1dalfovo1999theory} (here we ignore the initial phase $\theta_{m_F}^{(j)}$, which is independent of $\Delta \omega$) , where $m$ is the atomic mass, $\hbar$ is the reduced Planck constant,  ${y_{m_F}^{(j)}}$ is the position of the atomic cloud in the $\left\vert m_F\right\rangle$ state of the $j$ path after the TOF, and ${\upsilon_{m_F}^{(j)}}$ is the velocity of the atomic cloud in the $\left\vert F,m_{F}^{({j})}\right\rangle$ state. Here the $\left|m_F=2 \right\rangle$ state is chosen as the phase reference, and then we have $\phi_{m_F}^{(j)}=\varphi_{m_F}^{(j)}-\varphi_{2}^{(j)}$. We put $\phi_{m_F}^{(j)}$ into Eq.~\eqref{eqn:phi} and adopt $m_F=2$, and then we have 
	\begin{align}\label{eqn:phi-1}
		\Delta\phi ={\frac {m}{\hbar T_{\textit{TOF}}}}{\delta y_{1}}{\delta y_{2\leftarrow 1}^{(j)}}+{\frac {m}{\hbar }}{\delta v_{1}}{\delta y_{2\leftarrow 1}^{(j)}}+
		{\frac {m}{\hbar }}{\delta y_{1}}{\delta v_{2\leftarrow 1}^{(j)}}
	\end{align}
	where
	\begin{align}
		\label{eqn:spe1}&\delta v_{1}={v_{1}^{(I)}}-{v_{1}^{(II)}}=a{T_d}\\
		\label{eqn:spe2}&\delta v_{2\leftarrow 1}^{(j)}={v_{1}^{(j)}}-{v_{2}^{(j)}}= a{T_N}\\
		\label{eqn:pos1}&\delta y_{1}={y_{1}^{(I)}}-{y_{1}^{(II)}}\simeq a{T_d}T_{\textit{TOF}}\\
		\label{eqn:pos2}&\delta y_{2\leftarrow 1}^{(j)}={y_{1}^{(j)}}-{y_{2}^{(j)}}\simeq a{T_N}{T_{\textit{TOF}}}
	\end{align}
Here $\delta v_{1}$ and $\delta y_{1}$ are the velocity difference and position difference after TOF between the atomic wave packets of the same mode $\langle m_F=1 \rangle$ in two paths. $\delta v_{2\leftarrow 1}^{(j)}$ and $\delta y_{2\leftarrow 1}^{(j)}$ are the velocity difference and position difference after TOF of the adjacent atomic wave packets in the same path, respectively. $a$ is the acceleration between two adjacent modes in the trap, which is derived from the interaction between the adjacent states and the difference of gradient magnetic forces between them. We get $a\simeq \SI[per-mode=symbol]{7.3}{\meter\per\square\second}$ from experimental measurements in a duration of tens of microseconds\cite{7.1PhysRevATPJ}.  
	
Then, by substituting Eqs.~\eqref{eqn:spe1}-\eqref{eqn:pos2} into Eq.~\eqref{eqn:phi-1}, we have the relative phase of adjacent component fringes
\begin{align} \label{eqn:phi-2}
	\Delta\phi \simeq 3\frac{m}{\hbar }a^{2}{T_d}T_{\textit{TOF}}{T_N}=\Delta \omega {T_N}
\end{align}
where $\Delta \omega = 3\frac{m}{\hbar }a^{2}{T_d}T_{\textit{TOF}}$ is the relative phase evolution rate. After considering the systematic error, we have
\begin{align}\label{eqn:theory}
	\Delta \omega \simeq 3\frac{m}{\hbar }a \delta y_1 \left[ 1\pm (\frac{2\Delta a}{a}+\frac{\Delta T_{d}}{T_{d}})\right]
\end{align}
Here $\Delta a = \SI[per-mode=symbol]{0.3}{\meter\per\square\second}$ comes from the position uncertainty of the wave packets when we measured the acceleration $a$ by \SI{26}{\ms} TOF image. $\delta y_1 = a{T_{d}}{T_{\textit{TOF}}}$ is the distance of the two wave packets after TOF and $\Delta T_d \simeq \SI{100}{\nano\second}$ is the uncertainty of $T_d$. Meanwhile, considering that the gradient magnetic field keeps on during the rf duration($\tau_R=\SI{10.0}{\micro\second}$), we actually estimate $\Delta \omega$ with $T_d+\tau_R/2$ instead of $T_d$, where $\tau_R/2$ is a compensation time.
\end{subappendices}

\bibliographystyle{apsrev}
\bibliography{citelist5}
	
\end{document}